\documentclass[aip,apl,twocolumn,reprint,nofootinbib]{revtex4-1}
\usepackage[hidelinks]{hyperref}
\usepackage{graphicx}
\usepackage{amsmath, amssymb,upgreek}
\usepackage{xcolor}
\usepackage[normalem]{ulem}
\hypersetup{
  colorlinks   = true, 
  urlcolor     = blue, 
  linkcolor    = blue, 
  citecolor   = blue 
 }

\begin{document}

\title{A capacitance spectroscopy-based platform for realizing gate-defined electronic lattices}
\author{T. Hensgens}
\thanks{These authors contributed equally to this work}
\author{U. Mukhopadhyay}
\thanks{These authors contributed equally to this work}
\author{P. Barthelemy}
\thanks{These authors contributed equally to this work}
\author{R.~F.~L. Vermeulen}
\author{R.~N. Schouten}
\affiliation{QuTech and Kavli Institute of Nanoscience, Delft University of Technology, 2600 GA Delft, The Netherlands}
\author{S. Fallahi}
\author{G.C. Gardner}
\affiliation{Department of Physics and Astronomy, and Station Q Purdue, Purdue University, West Lafayette, Indiana 47907, USA}
\author{C. Reichl}
\author{W. Wegscheider}
\affiliation{Solid State Physics Laboratory, ETH Z\"{u}rich, 8093 Z\"{u}rich, Switzerland}
\author{M.~J. Manfra}
\affiliation{Department of Physics and Astronomy, and Station Q Purdue, Purdue University, West Lafayette, Indiana 47907, USA}
\author{L.~M.~K. Vandersypen}
\thanks{Correspondence should be sent to l.m.k.vandersypen@tudelft.nl}
\affiliation{QuTech and Kavli Institute of Nanoscience, Delft University of Technology, 2600 GA Delft, The Netherlands}
\date{\today}

\begin{abstract}
Electrostatic confinement in semiconductors provides a flexible platform for the emulation of interacting electrons in a two-dimensional lattice, including in the presence of gauge fields. This combination offers the potential to realize a wide host of quantum phases.
Capacitance spectroscopy provides a technique that allows to directly probe the density of states of such two-dimensional electron systems. 
Here we present a measurement and fabrication scheme that builds on capacitance spectroscopy and allows for the independent control of density and periodic potential strength imposed on a two-dimensional electron gas.
We characterize disorder levels and (in)homogeneity and develop and optimize different gating strategies at length scales where interactions are expected to be strong.
A continuation of these ideas might see to fruition the emulation of interaction-driven Mott transitions or Hofstadter butterfly physics.
\end{abstract}

\pacs{}
\maketitle

\section{Introduction}

Artificial lattice structures have the potential for realizing a host of distinct quantum phases\cite{Cirac2012}. Of these, the inherent length scale of optical platforms allows for a clean emulation of quantum mechanical band physics, but also means interactions are weak and going beyond a single-particle picture is difficult\cite{Rechtsman2013,Tanese2014}. For electronic implementations in solid-state, interactions can be made non-perturbatively strong, potentially leading to a host of emergent phenomena. An example is shown in graphene superlattices, where not only Hofstadter’s butterfly physics\cite{Hofstadter1976,Dean2013,Ponomarenko2013,Hunt2013} but also interaction-driven and emergent fractional quantum Hall states in the butterfly appear\cite{Yu2014}. The ideal platform would host a designer lattice with tunable electron density and lattice strength, allowing to emulate band physics for a wide variety of lattice types and giving access to the strong-interaction limit of correlated Mott phases\cite{Stafford1994,Manousakis2002,Byrnes2008,Singha2011,Barthelemy2013}. Semiconductor heterostructures with nano-fabricated gate structures provide this flexibility in lattice design and operation, yet inherent disorder in the host materials as well as the short length scales required make the realization of clean lattices difficult\cite{Ensslin1990,Geisler2004,Albrecht2001}.

In this Letter, we introduce a novel experimental platform for realizing artificial gate-induced lattices in semiconductors, based on a capacitance spectroscopy technique \cite{Ashoori1991,Ashoori1992,Ashoori1993}, with the potential to observe both single-particle band structure physics such as Hofstadter's butterfly and many-body physics such as the interaction driven Mott insulator transition. We discuss different gating strategies for imprinting a two-dimensional periodic potential at length scales where interactions are expected to be strong, characterize intrinsic disorder levels and show first measurements of double gate devices.

\section{Heterostructure and Capacitance spectroscopy}
\begin{figure}[!htb]
	\centering
	\includegraphics{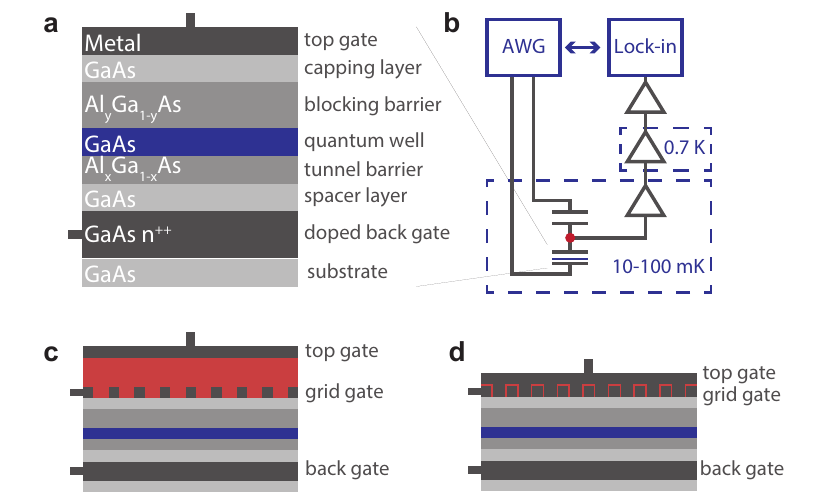} 
	\caption{(a) Schematic diagram showing the various layers of the samples with a single global gate.  
	(b) Bridge set-up for equilibrium capacitance measurements, where sinusoidal signals are applied by a waveform generator (WG) on both the sample back gate and on a reference capacitor of 45 pF. The relative amplitude and phase difference between these two signals are adjusted to maintain a constant zero voltage at the bridge point (red dot), which is amplified in different stages and read out using a lock-in amplifier. The bridge point is connected to the grid gate when there is a grid gate present, and to the top gate otherwise.
    (c)-(d) Schematic diagrams of two different two-layer gate geometries, designed to impose a periodic potential on the 2DEG, comprising either of a deposited dielectric (c) or a dielectric obtained by oxidation of the first metallic layer (d). Dielectric spacer is depicted in red. The other colors are as in panel (a).}
	\label{fig:fig1}
\end{figure}

To host the 2D electron gas (2DEG), we use a GaAs quantum well with AlGaAs barriers, grown by molecular beam epitaxy. The substrate contains a highly Si-doped GaAs layer that acts as a back gate. It is tunnel coupled to the 2DEG through a Al$_x$Ga$_{1-x}$As tunnel barrier (see Fig.~\ref{fig:fig1}a and Table \ref{table:I}). There is no doping layer above the quantum well in order to avoid an important source of disorder. A metallic top gate is fabricated on the surface. A variable capacitor forms between the back and top gates: when an alternating potential difference is applied between them, electrons tunnel back and forth between the back gate and the 2DEG, modifying the capacitance by an amount proportional to the density of states (DOS) of the 2DEG. The tunnel frequency depends mainly on the thickness and the Al content ($x$) of the tunnel barrier.
At the limits of zero or infinite DOS, the system behaves like a simple parallel plate capacitor, described by the distance between top gate and back gate or top gate and 2DEG, respectively.
The capacitance is read out using a bridge design with a reference capacitor \cite{Cavicchi1988}, where the voltage at the bridge point is kept constant (Fig.~\ref{fig:fig1}b) by changing the amplitude ratio and phase difference of AC signals applied to each capacitor (see supplementary material section A for experimental details).

To impose a periodic potential in the 2DEG, we pattern a metallic gate into a grid shape before making the top gate.
From a capacitance spectroscopy perspective, this double-gate structure can be made with two different designs.
In the first design, the top gate is separated from the grid gate by a thick dielectric layer, rendering its capacitance to the grid gate negligible (a few pF compared to tens of pF). In that case, we can ignore the grid gate from an AC perspective altogether (Fig.~\ref{fig:fig1}c).
Alternatively, we can minimize the separation between the two gate layers, such that the capacitance between the two top gates (100's of pF) exceeds the sample capacitance. Here the two gates effectively form a single gate (Fig.~\ref{fig:fig1}d), as seen in AC.
We investigate both designs below, starting with describing the fabrication (limits) and following with measurements of disorder levels and imposed potentials.

\section{Gate Design and Fabrication}

\begin{figure}[!htb]
	\centering
	\includegraphics{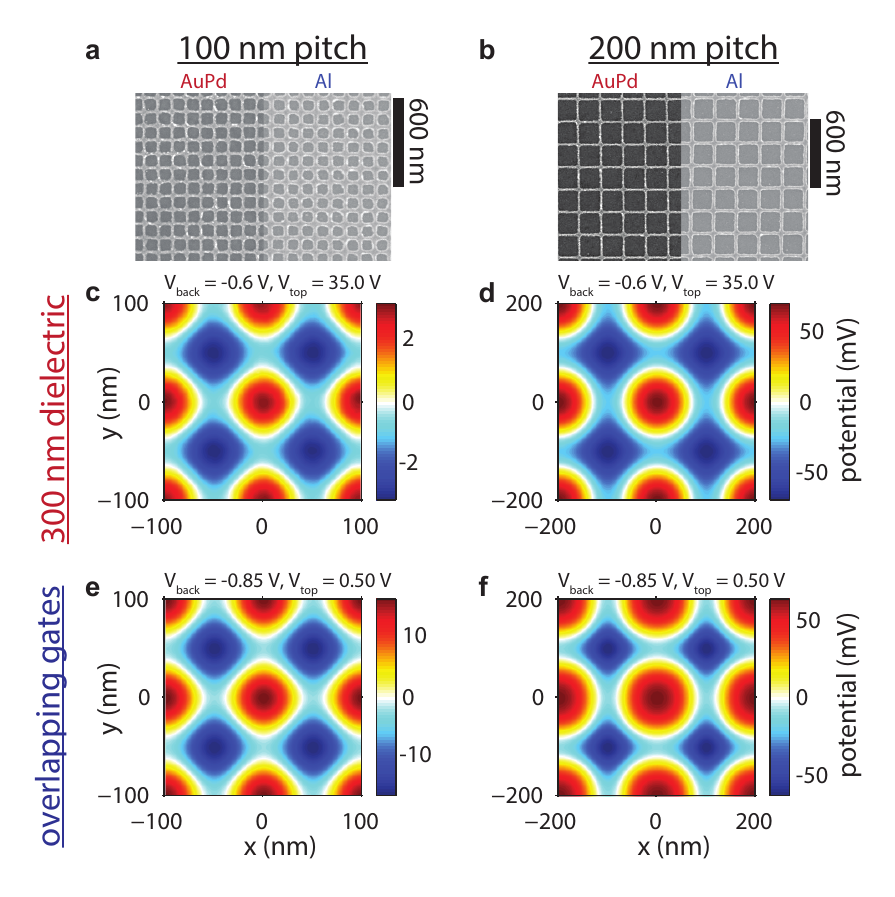}
	\caption{(a) Electron micrograph of 100 nm periodic AuPd and Al grid gate structures for two different gate designs.
    (b) Similar, for 200 nm periodic gate structures.
    (c)-(f) Electrostatic simulations of imparted potential in the 2DEG in both designs and both gate pitches (100 and 200 nm) using denoted gate voltages. For (c)-(d), we use a 350 nm SiO$_2$ dielectric and flat top gate. For (e)-(f), we use a 5 nm spacer dielectric (oxidized Aluminum oxide) separating the two top gates. Voltages used are roughly the empirical maximum voltage difference we can set for both designs (see fig \ref{fig:periodic}), $V_{\mathrm{grid}}=-0.45$ V for both. Width of the metal grids are taken as 22 nm and 25 nm for AuPd and Al grids respectively for reasons explained below.}\label{fig:simulation}
\end{figure}

We distinguish devices with a single global gate (Fig.~\ref{fig:fig1}a) and devices with two layers of gates: a grid gate and a uniform global gate on top (Fig.~\ref{fig:fig1}c-d). The former will be used to characterize disorder levels in the next section, whereas the latter allows for the imposition of a periodic potential. The strength of the imparted periodic potential depends on the dielectric choice (thick or thin, compare Fig.~\ref{fig:fig1}c,d), gate design, grid gate pitch and the maximum voltages that can be applied.  Grid gates are made with a pitch of 100 - 200 nm (Fig.~\ref{fig:simulation}a-b), which is mainly limited by the fabrication constraints. The maximum voltage is determined by the onset of leakage through the heterostructure or the accumulation of charges in the capping layer, and thus depends on heterostructure details such as the Al concentration and layer thicknesses.

The expected imparted potentials at the 2DEG with typical maximum voltages for both designs are shown in Fig.~\ref{fig:simulation}(c-f) (calculated using COMSOL electrostatic simulation software). In order to observe a Mott transition and the corresponding localization of electrons on individual sites, the periodic potential amplitude must exceed the local Coulomb repulsion (typically several meVs) \cite{Kouwenhoven1997}.
For 200 nm grids, both designs show similar maximum effective periodic potentials, and they should suffice for the formation of quantum dots. For the 100 nm grids, however, the achievable potentials exceed the charging energy only when using the overlapping gate design.
For the smaller pitch grid, effective shielding of the top gate voltage by the grid gate is larger when the top gate is farther away from the heterostructure. Therefore, an overlapping gate design is required to go to sufficiently strong periodic potentials for localization at 100 nm site-to-site pitch.

Furthermore, we note that screening induced by mobile charges in the back gate region has both desired and undesirable consequences. An intended benefit is that disorder from charged impurities or defects in the heterostructure is partly screened, and the more so the closer to the back gate the impurities or defects are located \cite{Barthelemy2013}. However, electron-electron interactions and the gate-voltage imposed potential modulation itself are partly screened as well, and more so as the lattice dimension is reduced. 

Double gate devices with either a thick (Fig.~\ref{fig:fig1}c) or a thin dielectric (Fig.~\ref{fig:fig1}d) between the two gates require different fabrication processes.
Here we discuss the fabrication of the active regions in both designs, which have a size of 200 $\upmu$m by 200 $\upmu$m. The detailed information for all steps in the fabrication is provided in the supplementary material section B.
In both designs, the square grid metallic gates are fabricated at pitches of 100-200 nm using electron beam lithography and evaporation of metals in a standard lift-off process (Fig.~\ref{fig:simulation}a-b).
In the first design, both gates are made of Ti/Au(Pd) and separated by $>$ 200 nm layer of oxide, such as plasma-enhanced chemical vapor deposition grown SiO$_x$ or plasma-enhanced atomic layer deposition grown AlO$_x$.
In the second design, both gates are made of Al, and an oxygen (remote) plasma oxidation step is used after depositing the first Al layer to ensure sufficient electrical isolation between the two layers by transforming part of the Al gate to Aluminum oxide \cite{Angus2007}. In this design, we measure resistances exceeding 1 G$\Omega$ over several V.

\begin{figure}[!htb]
	\centering
	\includegraphics{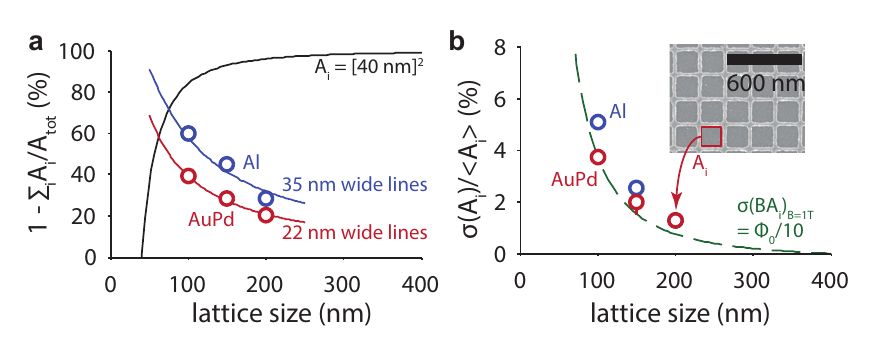}
	\caption{(a) Fraction of surface area covered by the grid gate as function of lattice size. Black line indicates a grid with the smallest possible plaquettes allowed by the lift-off process, whereas the blue (red) line indicates the percentage of surface area covered by a grid with metallic lines of 35 nm (22 nm). (b) Variation in relative area of non-metal plaquettes in the grid gate layer ($A_i$, see inset) as function of lattice size, as a measure of fabrication (in)homogeneity.  The green dashed line indicates variations in plaquette area that coincide with variations of a tenth of a flux quantum at 1 T (see Discussion below). Blue (red) points indicate grid gates made of Al (AuPd) for both figures.}
	\label{fig:limits}
\end{figure}

Because of the fabrication process, there are limits in the periodicity and homogeneity of the grid gate layer. We typically find (1) that plaquettes of smaller size than 40 nm x 40 nm will not lift off and that (2) the grain size of a particular metal determines the narrowest lines that can be made reliably with liftoff. For the materials used here, AuPd and Al, these effects limit the minimum lattice pitch (Fig.~\ref{fig:limits}a).
Furthermore, we have analyzed the homogeneity of the lattices by using image processing techniques to give the statistics of the non-metal plaquette areas (Fig.~\ref{fig:limits}b). A more relaxed lattice constant means higher relative homogeneity
but this is not necessarily helpful: it also increases the flux through a single plaquette when a perpendicular magnetic field is applied (relevant for Hofstadter butterfly physics, as will be described below) and it decreases the charging energy, relevant for Mott interaction physics.

\section{Measurements}

\subsection{Global gates: disorder levels}

\begin{figure}[!htb]
	\centering
	\includegraphics{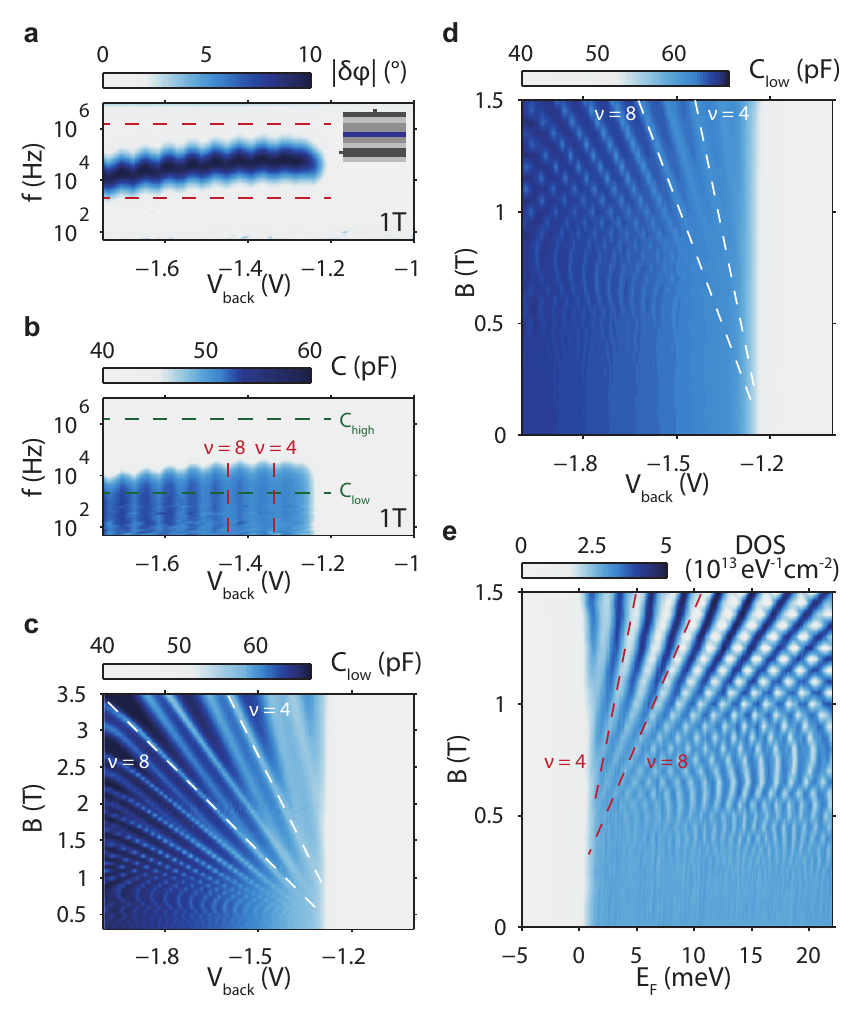}
	\caption{(a) Bridge equilibrium phase as function of back gate bias and measurement frequency.
    (b) Global gate capacitance as function of back gate bias and measurement frequency.
    (c) Landau fan diagram: device capacitance as function of back gate bias and magnetic field, showing onset of accumulation, integer quantum Hall levels and exchange splitting.
    (d) Zoom in of Landau fan diagram for low field regime of (c).
    (e) Calculated density of states (DOS) from (d), allowing us to assess disorder from Landau level visibility.
    The gaps at filling factors $\nu=4$ and $\nu=8$ are indicated. At lower fields, the small Landau level spacing leads to aliasing in the image.}
	\label{fig:disorder}
\end{figure}

In order to assess disorder levels, we first measure the devices with a single uniform top gate. We measure the capacitance at frequencies below and above the rate at which electrons tunnel between the 2DEG and the doped back gate region as a function of bias voltage (Fig.~\ref{fig:disorder}a-b) and magnetic field.
Having measured the capacitance at low and high frequencies, we calculate the equilibrium DOS. There are essentially two unknown parameters in this conversion, namely the distance from top to bottom gate and the relative location of the 2DEG itself.
The former can be directly inferred from the capacitance at high frequency, the latter by using either the known effective mass or the Landau level splitting with magnetic field as benchmarks (see supplementary material section C for details on this conversion).

As a magnetic field is turned on, we see the onset of Landau level formation. For magnetic fields above 2 T, we observe a splitting between the spin subbands of the Landau levels which increases with the applied magnetic field (Fig.~\ref{fig:disorder}c). For a given magnetic field, the separation between the two subbands of any Landau level is significantly larger than the Zeeman energy with g = -0.44 for bulk electrons in GaAs \cite{Ashoori1991}. This enhanced Zeeman splitting is an effect of the Coulomb repulsion between electrons in the same subband \cite{Ando1974}.

We focus on the low-field data (Fig.~\ref{fig:disorder}d) and infer disorder levels from the density of states data (Fig.~\ref{fig:disorder}e).
Gaussian fits to the Landau levels yield typical widths ranging between 0.4-1 meV at densities above 10$^{11}$ cm$^{-2}$, which, although hard to compare directly to the mobilities reported for transport-based wafers \cite{Ensslin1990,Geisler2004,Albrecht2001}, is comparable to previously reported values for similar heterostructures\cite{Jang2016}.
The Landau levels themselves (aliased at low fields in Fig.~\ref{fig:disorder}d) become visible above fields of roughly 0.25 T, corresponding to densities per Landau level of 1.2$\times$10$^{10}$ cm$^{-2}$ and cyclotron gaps of 0.43 meV.
The Landau level width did not change when we increased the mixing chamber temperature from 10 mK to 100 mK or when we varied the excitation voltage. Furthermore, the Landau level width was consistent across fabrication schemes, but did vary with the wafer used. Therefore, we consider it a heuristic metric for the achievable disorder levels on a particular wafer.

We have tried to optimize wafer design to minimize this disorder, whilst allowing for the imposition of a periodic potential. All in all, over twenty different GaAs/Al$_x$Ga$_{1-x}$As wafers grown by molecular beam epitaxy have been used. Growth details of the wafers can be found in Table \ref{table:I}.

The initial wafer (W1) design was based on Dial et. al.\cite{Dial2007}, and was grown on a conducting substrate.
This simplifies the fabrication of single-gate devices, as an unpatterned ohmic back gate contact can be directly evaporated on the back side of the wafer, while simple metallic pads fabricated on the front side can be directly bonded to and used as a top gate.
A double-gate design requires dedicated bond pads, which would give a sizable contribution to the total capacitance when fabricated directly on the wafer.
The device used for Fig.~\ref{fig:periodic}a-b in the main text, fabricated on one of the first rounds of wafers (W2), therefore, had bond pads on top of the thick dielectric separating the two gates.
This strategy is not compatible with the second design, where there is no thick dielectric layer, and also gives a very low wire bonding yield due to poor adhesion of the dielectric layers on the GaAs surface. Furthermore, handling both sides of a substrate during fabrication risks contaminating the front surface, and is particularly suboptimal when detailed features (grid gates) are present as well.
Subsequent wafers were therefore grown with a 400-800 nm thin degenerately Si doped back gate region that is contacted from the front side of the wafer, and is etched to form electrically isolated device and bond pad mesas.\\
We have further tried to optimize the wafer stacks aiming to increase the amplitude of the periodic potential at the 2DEG and to decrease disorder levels. A stronger periodic potential can be obtained by either increasing the maximum possible gate voltage, reducing the separation between the grid gate and the 2DEG or increasing the distance between the 2DEG and the back gate. The latter may also reduce disorder caused by dopant diffusion from the back gate. Increasing the quantum well thickness is also expected to reduce the effect of disorder by accommodating more of the electron wave-function away from the interfaces.
Concretely, we have first varied spacer layer thickness (25 and 35 nm) and quantum well widths (15 and 30 nm).
In further attempts to optimize the trade-off between the periodic potential that can be set at a fixed voltage and the maximum voltage we can apply to the gates before leakage sets in, we varied the blocking barrier thickness (40, 50, 60 and 70 nm) and fabricated devices with a thin dielectric layer (see wafers M1 and W3) added underneath the grid gate.
None of these, however, managed to noticeably increase the maximum potential we could impose on the 2DEG, or to decrease disorder levels. The strongest effect on disorder was obtained by changing the aluminum concentration in the Al$_x$Ga$_{1-x}$As blocking and tunnel barrier (from $x=0.31$ everywhere to $x=0.36$ in the blocking barrier and $x=0.20$ in the tunnel barrier), while slightly increasing the tunnel barrier thickness in order to keep the tunnel rates roughly the same (see Table~\ref{table:I}). The measurements shown in Fig.~\ref{fig:disorder} and Fig.s.~\ref{fig:periodic}c-d are taken on this optimized wafer, called M2. 

\begin{table*}[!htb]
\centering
\caption{Heterostructure details.}\label{table:I}
\begin{tabular}{llllll} \hline
 & W1 & W2 & M1 & W3 & M2 \\ \hline
capping layer & GaAs & GaAs & GaAs & GaAs & GaAs \\
 & 10 nm & 10 nm & 5 nm & 10 nm & 5 nm \\
blocking barrier (Al content) & 0.316 & 0.316 & 0.316 & 0.315 & 0.360 \\
 & 60 nm & 60 nm & 40 nm & 60 nm & 60 nm \\
quantum well & GaAs & GaAs & GaAs & GaAs & GaAs \\
 & 23 nm & 23 nm & 23 nm & 23 nm & 23 nm \\
tunnel barrier (Al content) & 0.316 & 0.316 & 0.316 & 0.315 & 0.199 \\
 & 13 nm & 13 nm & 14 nm & 14 nm & 16 nm \\
spacer layer & GaAs & GaAs & GaAs & GaAs & GaAs \\
 & 25 nm & 15 nm & 15 nm & 15 nm & 15 nm \\
back gate & GaAs n$^{\mathrm{++}}$ & GaAs n$^{\mathrm{++}}$ & GaAs n$^{\mathrm{++}}$ & GaAs n$^{\mathrm{++}}$ & GaAs n$^{\mathrm{++}}$ \\
 & 800 nm & 800 nm & 400 nm & 400 nm & 400 nm \\ \hline
 tunneling frequency & 1 MHz & 200 kHz & 2 kHz & 30 kHz & 100 kHz \\
 at 0T, $n\approx 10^{11}$ cm$^{-2}$ & & & & & \\ \hline
 lowest field at which Landau & 3 T & 0.65 T & 0.50 T & 0.40 T & 0.25 T \\
 levels can be distinguished & (at 4 K) & & & & \\ \hline
 comments & n$^{\mathrm{++}}$ doped & n$^{\mathrm{++}}$ doped & & & \\
          & substrate     & substrate     & & & \\ \hline
\end{tabular}
\end{table*}

\subsection{Grid gates: periodic potential strength}

\begin{figure}[t]
	\centering
	\includegraphics{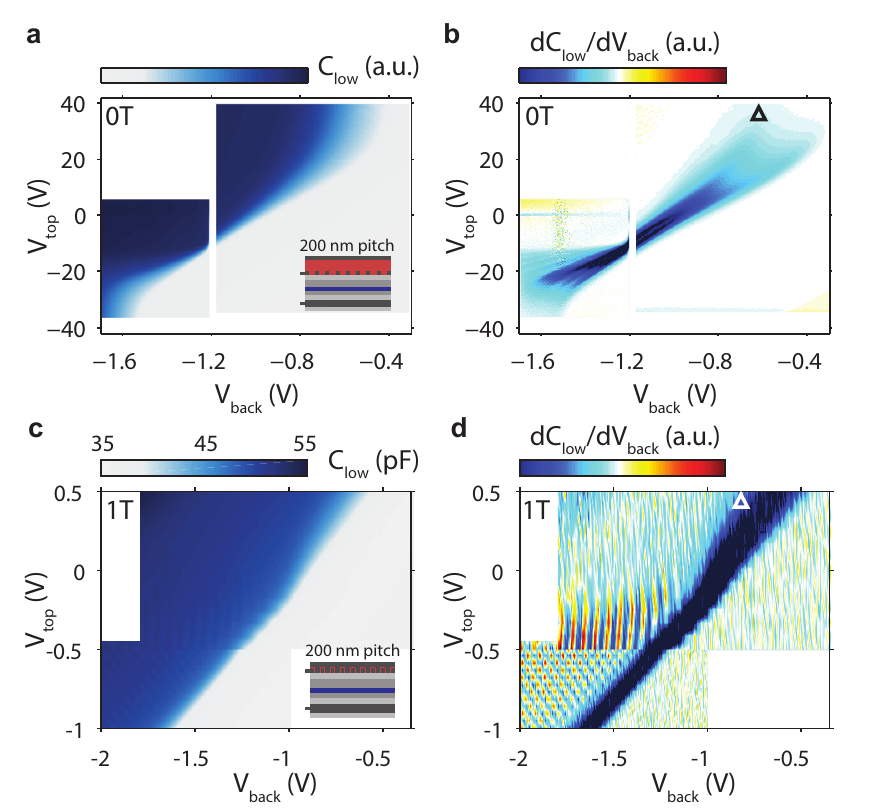}
	\caption{(a) Capacitance as function of back gate and top gate voltages for a device with a 200 nm periodic square grid gate and a 360 nm SiO2 dielectric separating the two gate layers (see inset and Fig. 3a). (b) Derivative of capacitance data. (c)-(d) Similar data taken for a device with aluminum overlapping gates (see inset) at 1 T. Black and white triangles in (b) and (d) indicate the gate voltages used in Fig.~\ref{fig:simulation} d and f respectively. The onset of accumulation shows broadening in (b) whereas Landau levels get blurred out with increasing top gate voltage in (d)}
	\label{fig:periodic}
\end{figure}

For measurements of two-layer gate devices of both designs (Fig.~\ref{fig:periodic}), we keep the grid gate potential fixed, given that it serves as the gate voltage of the first transistor in the amplification chain, and map out the remaining two gate voltages over as large a range as possible.
Initial devices of both designs indeed show accumulation as a function of the two gate voltages (transition from light grey to blue in Figs.~\ref{fig:periodic}a,c).
At voltages where we expect a flat periodic potential (close to the center of each panel in Fig.~\ref{fig:periodic}), and for our final set of devices, we can still distinguish well-defined Landau levels, indicating that the added fabrication steps themselves do not severely increase the disorder levels (data not shown). 
This disorder in the potential landscape also leads to a broadening of the onset of accumulation, seen in the center of  Figs.~\ref{fig:periodic}a,c.

For devices of the first design, this broadening increases as we move away from the center, along the grey-blue boundary  (Fig.~\ref{fig:periodic}a). This suggests that we see a gate-voltage induced spatial variation in the 2DEG potential that exceeds disorder levels (0.4-1 meV) at low densities. Based on electrostatic simulations of the strength of the imposed potential, the gate-voltage induced variation is indeed expected to exceed the disorder levels (Fig.~\ref{fig:simulation}).
The asymmetry between positive and negative top gate values seen in the data could possibly be explained by effective disorder levels being smaller when charges accumulate mainly underneath the grid gate, as compared to when charges accumulate mainly underneath the dielectric.
Finally, in Fig.~\ref{fig:periodic}b we resolve separate lines at the onset of accumulation for negative top gate voltages. Even though we expect to see evidence of miniband formation, we do not attribute these splittings to miniband formation, as they show a much larger periodicity in back gate voltage than the 6 mV expected from the density of states calculation (see below).

For devices of the second design, the widening of the onset of accumulation is less pronounced , but the effect of gating is seen at finite magnetic fields, where a voltage difference between the grid and top gate effectively blurs out the gaps between Landau levels (Fig.~\ref{fig:periodic}c-d). This indicates that the imposed local potential variation must be comparable to or stronger than the Landau level spacing at 1 T (1.7 meV). We conclude that also for the second design, the 200 nm periodic potential exceeds disorder levels.

Increasing further the amplitude of the potential variation induced by the gates was limited by saturation of the gating effect.
For the first gate design, we find a saturation to the effect of the top gate in gating the 2DEG at gate voltages exceeding 35 V in absolute value. This could be a sign of charges building up at interface of the capping layer and the dielectric, or in the dielectric itself, which screen the effect of the top gate. This saturation limits the potential we can impose on the 2DEG.
For the second gate design, a maximum voltage difference of roughly 2 V can be set between the back gate and the surface gates before leakage starts to occur.
As an attempt to allow for larger gate voltages before leakage through the heterostructure occurs, we have tried the same fabrication but with an additional 5 nm ALD-grown AlO$_x$ dielectric placed underneath the grid gate. This indeed prevents leakage but the gating effect saturated at the same voltages as where leakage occurred for devices without this additional dielectric. Therefore, 2 V was still the maximum voltage we could apply between the back and surface gates in the second design.

\section{Discussion: what to search for in future data}

As we have just seen, (i) the periodic potential exceeds disorder levels. In order to see Hofstadter's butterfly and Mott physics, however, we also need to (ii) be able to resolve the induced density of states modulations and (iii) the lattice potential from the grid itself should be sufficiently homogeneous. The latter two considerations will be discussed below, based on the data presented.

Using either gate design we find both gates to influence the accumulation of charges in the quantum well as expected, but neither shows clear evidence of a lattice potential imposed on the 2DEG (Fig.~\ref{fig:periodic}).
At zero magnetic field, a lattice potential would lead to minibands that manifest as periodic modulations in the density of states (and capacitance) with a period corresponding to two electrons per lattice site, or 5$\times$10$^{9}$ cm$^{-2}$ for a 200 nm square grid. Expressed in mV on the back gate, this corresponds to a period of 6 mV.
Furthermore at finite magnetic field, Landau levels are expected to show structure due to Hofstadter butterfly physics\cite{Pfannkuche1992,Geisler2004}, with the largest gaps expected around $k \pm 1/4$ of a flux quantum $\Phi_0$ threading each lattice plaquette (with $k$ an integer; $\Phi_0$ corresponds to 104 mT for a 200 nm grid).
Finally, a strong enough periodic potential would allow interaction effects to dominate. Miniband gaps are expected to split as filling starts to occur with a period of one electron per lattice site, akin to the interaction-driven Mott transition \cite{Byrnes2008}. 
None of these effects are visible in Fig.~\ref{fig:periodic} nor in many detailed targeted scans of magnetic field and gate voltages on devices with 200 and 100 nm grid gate periodicity.

If we compare the 5$\times$10$^{9}$ cm$^{-2}$ density modulations expected from miniband formation with the 1.2$\times$10$^{10}$ cm$^{-2}$ broadening of low-field Landau levels (global gate devices at high densities, i.e. we do not have evidence that we can resolve density variations below 1.2$\times$10$^{10}$ cm$^{-2}$), it is reasonable that gaps are not yet seen opening up at densities corresponding to the filling of (pairs of) electrons on each lattice site.
This suggests that either lattice size or wafer disorder has to be further reduced.
As it proves hard to lift off plaquettes of metal that are smaller than roughly 40 nm by 40 nm, there is not much room to reduce lattice dimension further in this particular fabrication scheme (Fig.~\ref{fig:limits}a).
For 100 nm pitch grids, the period of the density modulations is expected to be four times larger, but is still comparable to current best-case scenario Landau level broadening. 
As such, reducing intrinsic disorder seems necessary. An appropriate goal would be to make double layer gate devices with Landau levels that are distinguishable at fields below 100 mT.

The visibility of Hofstadter butterfly gaps depends not only on the intrinsic disorder in the device, but also on the inhomogeneity in the plaquette sizes, as this would entail a different number of flux quanta threading through different plaquettes.
If the size variations from electron micrographs of our devices translated to identical size variations in the periodic potential (Fig.~\ref{fig:limits}b), we should just be able to distinguish the largest gaps\cite{Geisler2004}.
It is hard to assess, however, whether this indicator from the electron micrographs directly correlates to the relevant physics in the 2DEG.

\section{Outlook}

There is room for further optimization of these devices.
On the heterostructure side, the distance between the back gate and the 2DEG can be further increased, compensating with a decreased Al content in the tunnel barrier to keep the tunnel rate fixed. Furthermore, part of the spacer layer can be grown at reduced temperatures, which has been shown to strongly reduce disorder by limiting the diffusing of Si dopants from the back gate region\cite{Jang2016}.
On the fabrication side, there is still room left for a modest reduction of the lattice periodicity with the current lift-off process. Even smaller length scales can be obtained by switching to dry etching of the grid pattern, albeit at an unknown impact to wafer disorder levels.

In summary, we have demonstrated a novel platform intended for the realization of artificial lattices of interacting particles.
Although fine tuning the design to the point where a sufficiently homogeneous and strong periodic potential can be applied remains to be done, the quantum Hall data already shows how the strong-interaction, low-temperature limit can be reached.
Such a platform has potential for studying the interaction-driven Mott insulator transition\cite{Imada1998,Byrnes2008} and Hofstadter butterfly physics\cite{Hofstadter1976} with finite interactions, and can be extended  from the steady-state measurements presented here to include time-domain measurements of excited states\cite{Dial2007}.

\begin{acknowledgments}
The authors acknowledge useful discussions with O.E.~Dial, R.C.~Ashoori, G.A.~Steele, R. Schmits, A.J.~Storm and the members of the Delft spin qubit team as well as experimental assistance from M.~Ammerlaan, J.~Haanstra, S.~Visser and R.~Roeleveld. This work is supported by the Netherlands Organization of Scientific Research (NWO) VICI program, the European Commission via the integrated project SIQS and the Swiss National Science Foundation. The work at Purdue was supported by the US Department of Energy, Office of Basic Energy Sciences, under Award number DE-SC0006671.  Additional support from the W. M. Keck Foundation and Microsoft Station Q is gratefully acknowledged.
\end{acknowledgments}

\bibliographystyle{apsrev4-1}
\bibliography{References}

\begin{thebibliography}{27}%
\makeatletter
\providecommand \@ifxundefined [1]{%
 \@ifx{#1\undefined}
}%
\providecommand \@ifnum [1]{%
 \ifnum #1\expandafter \@firstoftwo
 \else \expandafter \@secondoftwo
 \fi
}%
\providecommand \@ifx [1]{%
 \ifx #1\expandafter \@firstoftwo
 \else \expandafter \@secondoftwo
 \fi
}%
\providecommand \natexlab [1]{#1}%
\providecommand \enquote  [1]{``#1''}%
\providecommand \bibnamefont  [1]{#1}%
\providecommand \bibfnamefont [1]{#1}%
\providecommand \citenamefont [1]{#1}%
\providecommand \href@noop [0]{\@secondoftwo}%
\providecommand \href [0]{\begingroup \@sanitize@url \@href}%
\providecommand \@href[1]{\@@startlink{#1}\@@href}%
\providecommand \@@href[1]{\endgroup#1\@@endlink}%
\providecommand \@sanitize@url [0]{\catcode `\\12\catcode `\$12\catcode
  `\&12\catcode `\#12\catcode `\^12\catcode `\_12\catcode `\%12\relax}%
\providecommand \@@startlink[1]{}%
\providecommand \@@endlink[0]{}%
\providecommand \url  [0]{\begingroup\@sanitize@url \@url }%
\providecommand \@url [1]{\endgroup\@href {#1}{\urlprefix }}%
\providecommand \urlprefix  [0]{URL }%
\providecommand \Eprint [0]{\href }%
\providecommand \doibase [0]{http://dx.doi.org/}%
\providecommand \selectlanguage [0]{\@gobble}%
\providecommand \bibinfo  [0]{\@secondoftwo}%
\providecommand \bibfield  [0]{\@secondoftwo}%
\providecommand \translation [1]{[#1]}%
\providecommand \BibitemOpen [0]{}%
\providecommand \bibitemStop [0]{}%
\providecommand \bibitemNoStop [0]{.\EOS\space}%
\providecommand \EOS [0]{\spacefactor3000\relax}%
\providecommand \BibitemShut  [1]{\csname bibitem#1\endcsname}%
\let\auto@bib@innerbib\@empty
\bibitem [{\citenamefont {Cirac}\ and\ \citenamefont
  {Zoller}(2012)}]{Cirac2012}%
  \BibitemOpen
  \bibfield  {author} {\bibinfo {author} {\bibfnamefont {J.~I.}\ \bibnamefont
  {Cirac}}\ and\ \bibinfo {author} {\bibfnamefont {P.}~\bibnamefont {Zoller}},\
  }\href {\doibase 10.1038/nphys2275} {\bibfield  {journal} {\bibinfo
  {journal} {Nature Physics}\ }\textbf {\bibinfo {volume} {8}},\ \bibinfo
  {pages} {264} (\bibinfo {year} {2012})}\BibitemShut {NoStop}%
\bibitem [{\citenamefont {Rechtsman}\ \emph {et~al.}(2013)\citenamefont
  {Rechtsman}, \citenamefont {Zeuner}, \citenamefont {Plotnik}, \citenamefont
  {Lumer}, \citenamefont {Podolsky}, \citenamefont {Dreisow}, \citenamefont
  {Nolte}, \citenamefont {Segev},\ and\ \citenamefont
  {Szameit}}]{Rechtsman2013}%
  \BibitemOpen
  \bibfield  {author} {\bibinfo {author} {\bibfnamefont {M.~C.}\ \bibnamefont
  {Rechtsman}}, \bibinfo {author} {\bibfnamefont {J.~M.}\ \bibnamefont
  {Zeuner}}, \bibinfo {author} {\bibfnamefont {Y.}~\bibnamefont {Plotnik}},
  \bibinfo {author} {\bibfnamefont {Y.}~\bibnamefont {Lumer}}, \bibinfo
  {author} {\bibfnamefont {D.}~\bibnamefont {Podolsky}}, \bibinfo {author}
  {\bibfnamefont {F.}~\bibnamefont {Dreisow}}, \bibinfo {author} {\bibfnamefont
  {S.}~\bibnamefont {Nolte}}, \bibinfo {author} {\bibfnamefont
  {M.}~\bibnamefont {Segev}}, \ and\ \bibinfo {author} {\bibfnamefont
  {A.}~\bibnamefont {Szameit}},\ }\href {\doibase 10.1038/nature12066}
  {\bibfield  {journal} {\bibinfo  {journal} {Nature}\ }\textbf {\bibinfo
  {volume} {496}},\ \bibinfo {pages} {196} (\bibinfo {year}
  {2013})}\BibitemShut {NoStop}%
\bibitem [{\citenamefont {Tanese}\ \emph {et~al.}(2014)\citenamefont {Tanese},
  \citenamefont {Gurevich}, \citenamefont {Baboux}, \citenamefont {Jacqmin},
  \citenamefont {Lema{\^{i}}tre}, \citenamefont {Galopin}, \citenamefont
  {Sagnes}, \citenamefont {Amo}, \citenamefont {Bloch},\ and\ \citenamefont
  {Akkermans}}]{Tanese2014}%
  \BibitemOpen
  \bibfield  {author} {\bibinfo {author} {\bibfnamefont {D.}~\bibnamefont
  {Tanese}}, \bibinfo {author} {\bibfnamefont {E.}~\bibnamefont {Gurevich}},
  \bibinfo {author} {\bibfnamefont {F.}~\bibnamefont {Baboux}}, \bibinfo
  {author} {\bibfnamefont {T.}~\bibnamefont {Jacqmin}}, \bibinfo {author}
  {\bibfnamefont {A.}~\bibnamefont {Lema{\^{i}}tre}}, \bibinfo {author}
  {\bibfnamefont {E.}~\bibnamefont {Galopin}}, \bibinfo {author} {\bibfnamefont
  {I.}~\bibnamefont {Sagnes}}, \bibinfo {author} {\bibfnamefont
  {A.}~\bibnamefont {Amo}}, \bibinfo {author} {\bibfnamefont {J.}~\bibnamefont
  {Bloch}}, \ and\ \bibinfo {author} {\bibfnamefont {E.}~\bibnamefont
  {Akkermans}},\ }\href {\doibase 10.1103/PhysRevLett.112.146404} {\bibfield
  {journal} {\bibinfo  {journal} {Physical Review Letters}\ }\textbf {\bibinfo
  {volume} {112}},\ \bibinfo {pages} {146404} (\bibinfo {year}
  {2014})}\BibitemShut {NoStop}%
\bibitem [{\citenamefont {Hofstadter}(1976)}]{Hofstadter1976}%
  \BibitemOpen
  \bibfield  {author} {\bibinfo {author} {\bibfnamefont {D.}~\bibnamefont
  {Hofstadter}},\ }\href {\doibase 10.1103/PhysRevB.14.2239} {\bibfield
  {journal} {\bibinfo  {journal} {Physical Review B}\ }\textbf {\bibinfo
  {volume} {14}},\ \bibinfo {pages} {2239} (\bibinfo {year}
  {1976})}\BibitemShut {NoStop}%
\bibitem [{\citenamefont {Dean}\ \emph {et~al.}(2013)\citenamefont {Dean},
  \citenamefont {Wang}, \citenamefont {Maher}, \citenamefont {Forsythe},
  \citenamefont {Ghahari}, \citenamefont {Gao}, \citenamefont {Katoch},
  \citenamefont {Ishigami}, \citenamefont {Moon}, \citenamefont {Koshino},
  \citenamefont {Taniguchi}, \citenamefont {Watanabe}, \citenamefont {Shepard},
  \citenamefont {Hone},\ and\ \citenamefont {Kim}}]{Dean2013}%
  \BibitemOpen
  \bibfield  {author} {\bibinfo {author} {\bibfnamefont {C.~R.}\ \bibnamefont
  {Dean}}, \bibinfo {author} {\bibfnamefont {L.}~\bibnamefont {Wang}}, \bibinfo
  {author} {\bibfnamefont {P.}~\bibnamefont {Maher}}, \bibinfo {author}
  {\bibfnamefont {C.}~\bibnamefont {Forsythe}}, \bibinfo {author}
  {\bibfnamefont {F.}~\bibnamefont {Ghahari}}, \bibinfo {author} {\bibfnamefont
  {Y.}~\bibnamefont {Gao}}, \bibinfo {author} {\bibfnamefont {J.}~\bibnamefont
  {Katoch}}, \bibinfo {author} {\bibfnamefont {M.}~\bibnamefont {Ishigami}},
  \bibinfo {author} {\bibfnamefont {P.}~\bibnamefont {Moon}}, \bibinfo {author}
  {\bibfnamefont {M.}~\bibnamefont {Koshino}}, \bibinfo {author} {\bibfnamefont
  {T.}~\bibnamefont {Taniguchi}}, \bibinfo {author} {\bibfnamefont
  {K.}~\bibnamefont {Watanabe}}, \bibinfo {author} {\bibfnamefont {K.~L.}\
  \bibnamefont {Shepard}}, \bibinfo {author} {\bibfnamefont {J.}~\bibnamefont
  {Hone}}, \ and\ \bibinfo {author} {\bibfnamefont {P.}~\bibnamefont {Kim}},\
  }\href {\doibase 10.1038/nature12186} {\bibfield  {journal} {\bibinfo
  {journal} {Nature}\ }\textbf {\bibinfo {volume} {497}},\ \bibinfo {pages}
  {598} (\bibinfo {year} {2013})}\BibitemShut {NoStop}%
\bibitem [{\citenamefont {Ponomarenko}\ \emph {et~al.}(2013)\citenamefont
  {Ponomarenko}, \citenamefont {Gorbachev}, \citenamefont {Yu}, \citenamefont
  {Elias}, \citenamefont {Jalil}, \citenamefont {Patel}, \citenamefont
  {Mishchenko}, \citenamefont {Mayorov}, \citenamefont {Woods}, \citenamefont
  {Wallbank}, \citenamefont {Mucha-Kruczynski}, \citenamefont {Piot},
  \citenamefont {Potemski}, \citenamefont {Grigorieva}, \citenamefont
  {Novoselov}, \citenamefont {Guinea}, \citenamefont {Fal'ko},\ and\
  \citenamefont {Geim}}]{Ponomarenko2013}%
  \BibitemOpen
  \bibfield  {author} {\bibinfo {author} {\bibfnamefont {L.~A.}\ \bibnamefont
  {Ponomarenko}}, \bibinfo {author} {\bibfnamefont {R.~V.}\ \bibnamefont
  {Gorbachev}}, \bibinfo {author} {\bibfnamefont {G.~L.}\ \bibnamefont {Yu}},
  \bibinfo {author} {\bibfnamefont {D.~C.}\ \bibnamefont {Elias}}, \bibinfo
  {author} {\bibfnamefont {R.}~\bibnamefont {Jalil}}, \bibinfo {author}
  {\bibfnamefont {A.~A.}\ \bibnamefont {Patel}}, \bibinfo {author}
  {\bibfnamefont {A.}~\bibnamefont {Mishchenko}}, \bibinfo {author}
  {\bibfnamefont {A.~S.}\ \bibnamefont {Mayorov}}, \bibinfo {author}
  {\bibfnamefont {C.~R.}\ \bibnamefont {Woods}}, \bibinfo {author}
  {\bibfnamefont {J.~R.}\ \bibnamefont {Wallbank}}, \bibinfo {author}
  {\bibfnamefont {M.}~\bibnamefont {Mucha-Kruczynski}}, \bibinfo {author}
  {\bibfnamefont {B.~A.}\ \bibnamefont {Piot}}, \bibinfo {author}
  {\bibfnamefont {M.}~\bibnamefont {Potemski}}, \bibinfo {author}
  {\bibfnamefont {I.~V.}\ \bibnamefont {Grigorieva}}, \bibinfo {author}
  {\bibfnamefont {K.~S.}\ \bibnamefont {Novoselov}}, \bibinfo {author}
  {\bibfnamefont {F.}~\bibnamefont {Guinea}}, \bibinfo {author} {\bibfnamefont
  {V.~I.}\ \bibnamefont {Fal'ko}}, \ and\ \bibinfo {author} {\bibfnamefont
  {A.~K.}\ \bibnamefont {Geim}},\ }\href {\doibase 10.1038/nature12187}
  {\bibfield  {journal} {\bibinfo  {journal} {Nature}\ }\textbf {\bibinfo
  {volume} {497}},\ \bibinfo {pages} {594} (\bibinfo {year}
  {2013})}\BibitemShut {NoStop}%
\bibitem [{\citenamefont {Hunt}\ \emph {et~al.}(2013)\citenamefont {Hunt},
  \citenamefont {Sanchez-Yamagishi}, \citenamefont {Young}, \citenamefont
  {Yankowitz}, \citenamefont {LeRoy}, \citenamefont {Watanabe}, \citenamefont
  {Taniguchi}, \citenamefont {Moon}, \citenamefont {Koshino}, \citenamefont
  {Jarillo-Herrero},\ and\ \citenamefont {Ashoori}}]{Hunt2013}%
  \BibitemOpen
  \bibfield  {author} {\bibinfo {author} {\bibfnamefont {B.}~\bibnamefont
  {Hunt}}, \bibinfo {author} {\bibfnamefont {J.~D.}\ \bibnamefont
  {Sanchez-Yamagishi}}, \bibinfo {author} {\bibfnamefont {a.~F.}\ \bibnamefont
  {Young}}, \bibinfo {author} {\bibfnamefont {M.}~\bibnamefont {Yankowitz}},
  \bibinfo {author} {\bibfnamefont {B.~J.}\ \bibnamefont {LeRoy}}, \bibinfo
  {author} {\bibfnamefont {K.}~\bibnamefont {Watanabe}}, \bibinfo {author}
  {\bibfnamefont {T.}~\bibnamefont {Taniguchi}}, \bibinfo {author}
  {\bibfnamefont {P.}~\bibnamefont {Moon}}, \bibinfo {author} {\bibfnamefont
  {M.}~\bibnamefont {Koshino}}, \bibinfo {author} {\bibfnamefont
  {P.}~\bibnamefont {Jarillo-Herrero}}, \ and\ \bibinfo {author} {\bibfnamefont
  {R.~C.}\ \bibnamefont {Ashoori}},\ }\href {\doibase 10.1126/science.1237240}
  {\bibfield  {journal} {\bibinfo  {journal} {Science (New York, N.Y.)}\
  }\textbf {\bibinfo {volume} {340}},\ \bibinfo {pages} {1427} (\bibinfo {year}
  {2013})}\BibitemShut {NoStop}%
\bibitem [{\citenamefont {Yu}\ \emph {et~al.}(2014)\citenamefont {Yu},
  \citenamefont {Gorbachev}, \citenamefont {Tu}, \citenamefont {Kretinin},
  \citenamefont {Cao}, \citenamefont {Jalil}, \citenamefont {Withers},
  \citenamefont {Ponomarenko}, \citenamefont {Piot}, \citenamefont {Potemski},
  \citenamefont {Elias}, \citenamefont {Chen}, \citenamefont {Watanabe},
  \citenamefont {Taniguchi}, \citenamefont {Grigorieva}, \citenamefont
  {Novoselov}, \citenamefont {Fal'ko}, \citenamefont {Geim},\ and\
  \citenamefont {Mishchenko}}]{Yu2014}%
  \BibitemOpen
  \bibfield  {author} {\bibinfo {author} {\bibfnamefont {G.~L.}\ \bibnamefont
  {Yu}}, \bibinfo {author} {\bibfnamefont {R.~V.}\ \bibnamefont {Gorbachev}},
  \bibinfo {author} {\bibfnamefont {J.~S.}\ \bibnamefont {Tu}}, \bibinfo
  {author} {\bibfnamefont {A.~V.}\ \bibnamefont {Kretinin}}, \bibinfo {author}
  {\bibfnamefont {Y.}~\bibnamefont {Cao}}, \bibinfo {author} {\bibfnamefont
  {R.}~\bibnamefont {Jalil}}, \bibinfo {author} {\bibfnamefont
  {F.}~\bibnamefont {Withers}}, \bibinfo {author} {\bibfnamefont {L.~A.}\
  \bibnamefont {Ponomarenko}}, \bibinfo {author} {\bibfnamefont {B.~A.}\
  \bibnamefont {Piot}}, \bibinfo {author} {\bibfnamefont {M.}~\bibnamefont
  {Potemski}}, \bibinfo {author} {\bibfnamefont {D.~C.}\ \bibnamefont {Elias}},
  \bibinfo {author} {\bibfnamefont {X.}~\bibnamefont {Chen}}, \bibinfo {author}
  {\bibfnamefont {K.}~\bibnamefont {Watanabe}}, \bibinfo {author}
  {\bibfnamefont {T.}~\bibnamefont {Taniguchi}}, \bibinfo {author}
  {\bibfnamefont {I.~V.}\ \bibnamefont {Grigorieva}}, \bibinfo {author}
  {\bibfnamefont {K.~S.}\ \bibnamefont {Novoselov}}, \bibinfo {author}
  {\bibfnamefont {V.~I.}\ \bibnamefont {Fal'ko}}, \bibinfo {author}
  {\bibfnamefont {A.~K.}\ \bibnamefont {Geim}}, \ and\ \bibinfo {author}
  {\bibfnamefont {A.}~\bibnamefont {Mishchenko}},\ }\href {\doibase
  10.1038/nphys2979} {\bibfield  {journal} {\bibinfo  {journal} {Nature
  Physics}\ }\textbf {\bibinfo {volume} {10}},\ \bibinfo {pages} {525}
  (\bibinfo {year} {2014})}\BibitemShut {NoStop}%
\bibitem [{\citenamefont {Stafford}\ and\ \citenamefont {{Das
  Sarma}}(1994)}]{Stafford1994}%
  \BibitemOpen
  \bibfield  {author} {\bibinfo {author} {\bibfnamefont {C.~A.}\ \bibnamefont
  {Stafford}}\ and\ \bibinfo {author} {\bibfnamefont {S.}~\bibnamefont {{Das
  Sarma}}},\ }\href {\doibase 10.1103/PhysRevLett.72.3590} {\bibfield
  {journal} {\bibinfo  {journal} {Physical Review Letters}\ }\textbf {\bibinfo
  {volume} {72}},\ \bibinfo {pages} {3590} (\bibinfo {year}
  {1994})}\BibitemShut {NoStop}%
\bibitem [{\citenamefont {Manousakis}(2002)}]{Manousakis2002}%
  \BibitemOpen
  \bibfield  {author} {\bibinfo {author} {\bibfnamefont {E.}~\bibnamefont
  {Manousakis}},\ }\href {\doibase 10.1023/A:1014295416763} {\bibfield
  {journal} {\bibinfo  {journal} {Journal of Low Temperature Physics}\ }\textbf
  {\bibinfo {volume} {126}},\ \bibinfo {pages} {1501} (\bibinfo {year}
  {2002})}\BibitemShut {NoStop}%
\bibitem [{\citenamefont {Byrnes}\ \emph {et~al.}(2008)\citenamefont {Byrnes},
  \citenamefont {Kim}, \citenamefont {Kusudo},\ and\ \citenamefont
  {Yamamoto}}]{Byrnes2008}%
  \BibitemOpen
  \bibfield  {author} {\bibinfo {author} {\bibfnamefont {T.}~\bibnamefont
  {Byrnes}}, \bibinfo {author} {\bibfnamefont {N.}~\bibnamefont {Kim}},
  \bibinfo {author} {\bibfnamefont {K.}~\bibnamefont {Kusudo}}, \ and\ \bibinfo
  {author} {\bibfnamefont {Y.}~\bibnamefont {Yamamoto}},\ }\href {\doibase
  10.1103/PhysRevB.78.075320} {\bibfield  {journal} {\bibinfo  {journal}
  {Physical Review B}\ }\textbf {\bibinfo {volume} {78}},\ \bibinfo {pages}
  {075320} (\bibinfo {year} {2008})}\BibitemShut {NoStop}%
\bibitem [{\citenamefont {Singha}\ \emph {et~al.}(2011)\citenamefont {Singha},
  \citenamefont {Gibertini}, \citenamefont {Karmakar}, \citenamefont {Yuan},
  \citenamefont {Polini}, \citenamefont {Vignale}, \citenamefont {Katsnelson},
  \citenamefont {Pinczuk}, \citenamefont {Pfeiffer}, \citenamefont {West},\
  and\ \citenamefont {Pellegrini}}]{Singha2011}%
  \BibitemOpen
  \bibfield  {author} {\bibinfo {author} {\bibfnamefont {A.}~\bibnamefont
  {Singha}}, \bibinfo {author} {\bibfnamefont {M.}~\bibnamefont {Gibertini}},
  \bibinfo {author} {\bibfnamefont {B.}~\bibnamefont {Karmakar}}, \bibinfo
  {author} {\bibfnamefont {S.}~\bibnamefont {Yuan}}, \bibinfo {author}
  {\bibfnamefont {M.}~\bibnamefont {Polini}}, \bibinfo {author} {\bibfnamefont
  {G.}~\bibnamefont {Vignale}}, \bibinfo {author} {\bibfnamefont {M.~I.}\
  \bibnamefont {Katsnelson}}, \bibinfo {author} {\bibfnamefont
  {A.}~\bibnamefont {Pinczuk}}, \bibinfo {author} {\bibfnamefont {L.~N.}\
  \bibnamefont {Pfeiffer}}, \bibinfo {author} {\bibfnamefont {K.~W.}\
  \bibnamefont {West}}, \ and\ \bibinfo {author} {\bibfnamefont
  {V.}~\bibnamefont {Pellegrini}},\ }\href {\doibase 10.1126/science.1204333}
  {\bibfield  {journal} {\bibinfo  {journal} {Science (New York, N.Y.)}\
  }\textbf {\bibinfo {volume} {332}},\ \bibinfo {pages} {1176} (\bibinfo {year}
  {2011})}\BibitemShut {NoStop}%
\bibitem [{\citenamefont {Barthelemy}\ and\ \citenamefont
  {Vandersypen}(2013)}]{Barthelemy2013}%
  \BibitemOpen
  \bibfield  {author} {\bibinfo {author} {\bibfnamefont {P.}~\bibnamefont
  {Barthelemy}}\ and\ \bibinfo {author} {\bibfnamefont {L.~M.~K.}\ \bibnamefont
  {Vandersypen}},\ }\href {\doibase 10.1002/andp.201300124} {\bibfield
  {journal} {\bibinfo  {journal} {Annalen der Physik}\ }\textbf {\bibinfo
  {volume} {525}},\ \bibinfo {pages} {808} (\bibinfo {year}
  {2013})}\BibitemShut {NoStop}%
\bibitem [{\citenamefont {Ensslin}\ and\ \citenamefont
  {Petroff}(1990)}]{Ensslin1990}%
  \BibitemOpen
  \bibfield  {author} {\bibinfo {author} {\bibfnamefont {K.}~\bibnamefont
  {Ensslin}}\ and\ \bibinfo {author} {\bibfnamefont {P.}~\bibnamefont
  {Petroff}},\ }\href {\doibase 10.1103/PhysRevB.41.12307} {\bibfield
  {journal} {\bibinfo  {journal} {Physical Review B}\ }\textbf {\bibinfo
  {volume} {41}},\ \bibinfo {pages} {12307} (\bibinfo {year}
  {1990})}\BibitemShut {NoStop}%
\bibitem [{\citenamefont {Geisler}\ \emph {et~al.}(2004)\citenamefont
  {Geisler}, \citenamefont {Smet}, \citenamefont {Umansky}, \citenamefont {von
  Klitzing}, \citenamefont {Naundorf}, \citenamefont {Ketzmerick},\ and\
  \citenamefont {Schweizer}}]{Geisler2004}%
  \BibitemOpen
  \bibfield  {author} {\bibinfo {author} {\bibfnamefont {M.}~\bibnamefont
  {Geisler}}, \bibinfo {author} {\bibfnamefont {J.}~\bibnamefont {Smet}},
  \bibinfo {author} {\bibfnamefont {V.}~\bibnamefont {Umansky}}, \bibinfo
  {author} {\bibfnamefont {K.}~\bibnamefont {von Klitzing}}, \bibinfo {author}
  {\bibfnamefont {B.}~\bibnamefont {Naundorf}}, \bibinfo {author}
  {\bibfnamefont {R.}~\bibnamefont {Ketzmerick}}, \ and\ \bibinfo {author}
  {\bibfnamefont {H.}~\bibnamefont {Schweizer}},\ }\href {\doibase
  10.1103/PhysRevLett.92.256801} {\bibfield  {journal} {\bibinfo  {journal}
  {Physical Review Letters}\ }\textbf {\bibinfo {volume} {92}},\ \bibinfo
  {pages} {256801} (\bibinfo {year} {2004})}\BibitemShut {NoStop}%
\bibitem [{\citenamefont {Albrecht}\ \emph {et~al.}(2001)\citenamefont
  {Albrecht}, \citenamefont {Smet}, \citenamefont {von Klitzing}, \citenamefont
  {Weiss}, \citenamefont {Umansky},\ and\ \citenamefont
  {Schweizer}}]{Albrecht2001}%
  \BibitemOpen
  \bibfield  {author} {\bibinfo {author} {\bibfnamefont {C.}~\bibnamefont
  {Albrecht}}, \bibinfo {author} {\bibfnamefont {J.}~\bibnamefont {Smet}},
  \bibinfo {author} {\bibfnamefont {K.}~\bibnamefont {von Klitzing}}, \bibinfo
  {author} {\bibfnamefont {D.}~\bibnamefont {Weiss}}, \bibinfo {author}
  {\bibfnamefont {V.}~\bibnamefont {Umansky}}, \ and\ \bibinfo {author}
  {\bibfnamefont {H.}~\bibnamefont {Schweizer}},\ }\href {\doibase
  10.1103/PhysRevLett.86.147} {\bibfield  {journal} {\bibinfo  {journal}
  {Physical Review Letters}\ }\textbf {\bibinfo {volume} {86}},\ \bibinfo
  {pages} {147} (\bibinfo {year} {2001})}\BibitemShut {NoStop}%
\bibitem [{\citenamefont {Ashoori}()}]{Ashoori1991}%
  \BibitemOpen
  \bibfield  {author} {\bibinfo {author} {\bibfnamefont {R.~C.}\ \bibnamefont
  {Ashoori}},\ }\href {https://arxiv.org/abs/cond-mat/0607739} {\emph {\bibinfo
  {title} {{PhD thesis}}}}\BibitemShut {NoStop}%
\bibitem [{\citenamefont {Ashoori}\ \emph {et~al.}(1992)\citenamefont
  {Ashoori}, \citenamefont {Stormer}, \citenamefont {Weiner}, \citenamefont
  {Pfeiffer}, \citenamefont {Pearton}, \citenamefont {Baldwin},\ and\
  \citenamefont {West}}]{Ashoori1992}%
  \BibitemOpen
  \bibfield  {author} {\bibinfo {author} {\bibfnamefont {R.}~\bibnamefont
  {Ashoori}}, \bibinfo {author} {\bibfnamefont {H.}~\bibnamefont {Stormer}},
  \bibinfo {author} {\bibfnamefont {J.}~\bibnamefont {Weiner}}, \bibinfo
  {author} {\bibfnamefont {L.}~\bibnamefont {Pfeiffer}}, \bibinfo {author}
  {\bibfnamefont {S.}~\bibnamefont {Pearton}}, \bibinfo {author} {\bibfnamefont
  {K.}~\bibnamefont {Baldwin}}, \ and\ \bibinfo {author} {\bibfnamefont
  {K.}~\bibnamefont {West}},\ }\href {\doibase 10.1103/PhysRevLett.68.3088}
  {\bibfield  {journal} {\bibinfo  {journal} {Physical Review Letters}\
  }\textbf {\bibinfo {volume} {68}},\ \bibinfo {pages} {3088} (\bibinfo {year}
  {1992})}\BibitemShut {NoStop}%
\bibitem [{\citenamefont {Ashoori}\ \emph {et~al.}(1993)\citenamefont
  {Ashoori}, \citenamefont {Lebens}, \citenamefont {Bigelow},\ and\
  \citenamefont {Silsbee}}]{Ashoori1993}%
  \BibitemOpen
  \bibfield  {author} {\bibinfo {author} {\bibfnamefont {R.}~\bibnamefont
  {Ashoori}}, \bibinfo {author} {\bibfnamefont {J.}~\bibnamefont {Lebens}},
  \bibinfo {author} {\bibfnamefont {N.}~\bibnamefont {Bigelow}}, \ and\
  \bibinfo {author} {\bibfnamefont {R.}~\bibnamefont {Silsbee}},\ }\href
  {\doibase 10.1103/PhysRevB.48.4616} {\bibfield  {journal} {\bibinfo
  {journal} {Physical Review B}\ }\textbf {\bibinfo {volume} {48}},\ \bibinfo
  {pages} {4616} (\bibinfo {year} {1993})}\BibitemShut {NoStop}%
\bibitem [{\citenamefont {Cavicchi}\ and\ \citenamefont
  {Silsbee}(1988)}]{Cavicchi1988}%
  \BibitemOpen
  \bibfield  {author} {\bibinfo {author} {\bibfnamefont {R.~E.}\ \bibnamefont
  {Cavicchi}}\ and\ \bibinfo {author} {\bibfnamefont {R.~H.}\ \bibnamefont
  {Silsbee}},\ }\href {\doibase 10.1063/1.1140002} {\bibfield  {journal}
  {\bibinfo  {journal} {Review of Scientific Instruments}\ }\textbf {\bibinfo
  {volume} {59}},\ \bibinfo {pages} {176} (\bibinfo {year} {1988})},\ \Eprint
  {http://arxiv.org/abs/https://doi.org/10.1063/1.1140002}
  {https://doi.org/10.1063/1.1140002} \BibitemShut {NoStop}%
\bibitem [{\citenamefont {Kouwenhoven}\ \emph {et~al.}(1997)\citenamefont
  {Kouwenhoven}, \citenamefont {Marcus}, \citenamefont {McEuen}, \citenamefont
  {Tarucha}, \citenamefont {Westervelt},\ and\ \citenamefont
  {Wingreen}}]{Kouwenhoven1997}%
  \BibitemOpen
  \bibfield  {author} {\bibinfo {author} {\bibfnamefont {L.~P.}\ \bibnamefont
  {Kouwenhoven}}, \bibinfo {author} {\bibfnamefont {C.~M.}\ \bibnamefont
  {Marcus}}, \bibinfo {author} {\bibfnamefont {P.~L.}\ \bibnamefont {McEuen}},
  \bibinfo {author} {\bibfnamefont {S.}~\bibnamefont {Tarucha}}, \bibinfo
  {author} {\bibfnamefont {R.~M.}\ \bibnamefont {Westervelt}}, \ and\ \bibinfo
  {author} {\bibfnamefont {N.~S.}\ \bibnamefont {Wingreen}},\ }\enquote
  {\bibinfo {title} {Electron transport in quantum dots},}\ in\ \href {\doibase
  10.1007/978-94-015-8839-3_4} {\emph {\bibinfo {booktitle} {Mesoscopic
  Electron Transport}}},\ \bibinfo {editor} {edited by\ \bibinfo {editor}
  {\bibfnamefont {L.~L.}\ \bibnamefont {Sohn}}, \bibinfo {editor}
  {\bibfnamefont {L.~P.}\ \bibnamefont {Kouwenhoven}}, \ and\ \bibinfo {editor}
  {\bibfnamefont {G.}~\bibnamefont {Sch{\"o}n}}}\ (\bibinfo  {publisher}
  {Springer Netherlands},\ \bibinfo {address} {Dordrecht},\ \bibinfo {year}
  {1997})\ pp.\ \bibinfo {pages} {105--214}\BibitemShut {NoStop}%
\bibitem [{\citenamefont {Angus}\ \emph {et~al.}(2007)\citenamefont {Angus},
  \citenamefont {Ferguson}, \citenamefont {Dzurak},\ and\ \citenamefont
  {Clark}}]{Angus2007}%
  \BibitemOpen
  \bibfield  {author} {\bibinfo {author} {\bibfnamefont {S.~J.}\ \bibnamefont
  {Angus}}, \bibinfo {author} {\bibfnamefont {A.~J.}\ \bibnamefont {Ferguson}},
  \bibinfo {author} {\bibfnamefont {A.~S.}\ \bibnamefont {Dzurak}}, \ and\
  \bibinfo {author} {\bibfnamefont {R.~G.}\ \bibnamefont {Clark}},\ }\href
  {\doibase 10.1021/nl070949k} {\bibfield  {journal} {\bibinfo  {journal} {Nano
  Letters}\ }\textbf {\bibinfo {volume} {7}},\ \bibinfo {pages} {2051}
  (\bibinfo {year} {2007})},\ \bibinfo {note} {pMID: 17567176},\ \Eprint
  {http://arxiv.org/abs/https://doi.org/10.1021/nl070949k}
  {https://doi.org/10.1021/nl070949k} \BibitemShut {NoStop}%
\bibitem [{\citenamefont {Ando}\ and\ \citenamefont {Uemura}(1974)}]{Ando1974}%
  \BibitemOpen
  \bibfield  {author} {\bibinfo {author} {\bibfnamefont {T.}~\bibnamefont
  {Ando}}\ and\ \bibinfo {author} {\bibfnamefont {Y.}~\bibnamefont {Uemura}},\
  }\href {\doibase 10.1143/JPSJ.37.1044} {\bibfield  {journal} {\bibinfo
  {journal} {Journal of the Physical Society of Japan}\ }\textbf {\bibinfo
  {volume} {37}},\ \bibinfo {pages} {1044} (\bibinfo {year} {1974})},\ \Eprint
  {http://arxiv.org/abs/https://doi.org/10.1143/JPSJ.37.1044}
  {https://doi.org/10.1143/JPSJ.37.1044} \BibitemShut {NoStop}%
\bibitem [{\citenamefont {Jang}\ \emph {et~al.}(2016)\citenamefont {Jang},
  \citenamefont {Hunt}, \citenamefont {Pfeiffer}, \citenamefont {West},\ and\
  \citenamefont {Ashoori}}]{Jang2016}%
  \BibitemOpen
  \bibfield  {author} {\bibinfo {author} {\bibfnamefont {J.}~\bibnamefont
  {Jang}}, \bibinfo {author} {\bibfnamefont {B.~M.}\ \bibnamefont {Hunt}},
  \bibinfo {author} {\bibfnamefont {L.~N.}\ \bibnamefont {Pfeiffer}}, \bibinfo
  {author} {\bibfnamefont {K.~W.}\ \bibnamefont {West}}, \ and\ \bibinfo
  {author} {\bibfnamefont {R.~C.}\ \bibnamefont {Ashoori}},\ }\href {\doibase
  10.1038/nphys3979} {\bibfield  {journal} {\bibinfo  {journal} {Nature
  Physics}\ }\textbf {\bibinfo {volume} {1}},\ \bibinfo {pages} {1} (\bibinfo
  {year} {2016})}\BibitemShut {NoStop}%
\bibitem [{\citenamefont {Dial}\ \emph {et~al.}(2007)\citenamefont {Dial},
  \citenamefont {Ashoori}, \citenamefont {Pfeiffer},\ and\ \citenamefont
  {West}}]{Dial2007}%
  \BibitemOpen
  \bibfield  {author} {\bibinfo {author} {\bibfnamefont {O.~E.}\ \bibnamefont
  {Dial}}, \bibinfo {author} {\bibfnamefont {R.~C.}\ \bibnamefont {Ashoori}},
  \bibinfo {author} {\bibfnamefont {L.~N.}\ \bibnamefont {Pfeiffer}}, \ and\
  \bibinfo {author} {\bibfnamefont {K.~W.}\ \bibnamefont {West}},\ }\href
  {\doibase 10.1038/nature05982} {\bibfield  {journal} {\bibinfo  {journal}
  {Nature}\ }\textbf {\bibinfo {volume} {448}},\ \bibinfo {pages} {176}
  (\bibinfo {year} {2007})}\BibitemShut {NoStop}%
\bibitem [{\citenamefont {Pfannkuche}\ and\ \citenamefont
  {Gerhardts}(1992)}]{Pfannkuche1992}%
  \BibitemOpen
  \bibfield  {author} {\bibinfo {author} {\bibfnamefont {D.}~\bibnamefont
  {Pfannkuche}}\ and\ \bibinfo {author} {\bibfnamefont {R.}~\bibnamefont
  {Gerhardts}},\ }\href {\doibase 10.1103/PhysRevB.46.12606} {\bibfield
  {journal} {\bibinfo  {journal} {Physical Review B}\ }\textbf {\bibinfo
  {volume} {46}},\ \bibinfo {pages} {12606} (\bibinfo {year}
  {1992})}\BibitemShut {NoStop}%
\bibitem [{\citenamefont {Imada}\ \emph {et~al.}(1998)\citenamefont {Imada},
  \citenamefont {Fujimori},\ and\ \citenamefont {Tokura}}]{Imada1998}%
  \BibitemOpen
  \bibfield  {author} {\bibinfo {author} {\bibfnamefont {M.}~\bibnamefont
  {Imada}}, \bibinfo {author} {\bibfnamefont {A.}~\bibnamefont {Fujimori}}, \
  and\ \bibinfo {author} {\bibfnamefont {Y.}~\bibnamefont {Tokura}},\ }\href
  {\doibase 10.1103/RevModPhys.70.1039} {\bibfield  {journal} {\bibinfo
  {journal} {Reviews of Modern Physics}\ }\textbf {\bibinfo {volume} {70}},\
  \bibinfo {pages} {1039} (\bibinfo {year} {1998})}\BibitemShut {NoStop}%
\end{thebibliography}%

\clearpage
\newpage

\renewcommand{\figurename}{Figure~S}
\renewcommand{\tablename}{Table~S}
\renewcommand{\theequation}{S\arabic{equation}}
\renewcommand{\thesection}{\Roman{section}} 
\renewcommand{\thesubsection}{\Alph{subsection}}
\setcounter{figure}{0}
\pagenumbering{roman}

\maketitle

\begin{centering}
{\Large Supplementary Information for} \\ \vspace{0.2cm}
{\Large \textbf{Capacitance spectroscopy of gate-defined electronic lattices}}\\
\vspace{0.4cm}

{\normalsize T. Hensgens$^1$, U Mukhopadhyay$^1$, P. Barthelemy$^1$, S. Fallahi$^2$, G. C. Gardner$^2$, C. Reichl$^3$, W. Wegscheider$^3$, M. J. Manfra$^2$ and L. M. K. Vandersypen$^1$}\\  
\vspace{0.4cm}
\normalsize{$^{1}$QuTech and Kavli Institute of Nanoscience, TU Delft, 2600 GA Delft, The Netherlands}\\
\normalsize{$^{2}$Department of Physics and Astronomy, and Station Q Purdue, Purdue University, West Lafayette, Indiana 47907, USA}\\
\normalsize{$^{3}$Solid State Physics Laboratory, ETH Z\"{u}rich, 8093 Z\"{u}rich, Switzerland}\\

\end{centering}

\subsection{Capacitance bridge}

The capacitance bridge is built on a printed circuit board (PCB) that is mounted on the 10 mK mixing chamber stage of a dilution fridge and whose main components are the device, the reference capacitor and a high electron mobility transistor (HEMT, that serves as the first amplifier).
By mounting the HEMT orthogonal to the PCB surface, we can apply magnetic fields to the sample without influencing the amplification chain.
All D/C lines on the sample PCB have R/C filters on top of the filtering in the fridge. A 10 and 40 M$\Omega$ resistance is used to bias the bridge point and top gate in D/C, respectively, and a bias-tee is added to bias the back gate on top of the measurement signal.
The high frequency lines are not attenuated in the fridge, as we found this to lead to ground loop issues, but are instead attenuated on the PCB itself. Measurement excitations are simple sinusoidal signals that get attenuated to the µV level and are generated using a signal generator at room temperature.
The bridge point voltage is amplified further at 0.7 K and at room temperature and measured using a lock-in.

An iterative scheme is implemented to minimize the bridge point voltage by updating the amplitude ratio and phase difference of the two excitations as gate voltages and applied magnetic field are changed. The excitation on the sample side is kept constant and the excitation on the reference capacitor side is updated based on the secant method.
For this, we model the bridge as a linear system of complex variables: $Y =AX+B$, where $X$ is the reference signal, $Y$ is the output from the lock-in, and $A$ and $B$ are complex numbers.
Given two iterations with reference signals $X_i$ and $X_{i+1}$ and respective output values $Y_i$ and $Y_{i+1}$, $A$ and $B$ are calculated as well as $X_{i+2}=-B/A$, which is subsequently set and $Y_{i+2}$ measured.
As the first two iterations, we take the last set reference signal  as well as a point with a typically 1 \% higher amplitude and a tenth of a degree increased phase.
Convergence is reached when the amplitude difference between the last two reference signals drops below some pre-defined value, typically chosen to be several parts per thousand of the amplitude itself. 
The sample capacitance $C_\mathrm{sample}$ follows from the reference capacitor value $C_\mathrm{ref}$ and the applied amplitude ratio $R=\frac{V_\mathrm{ref}}{V_\mathrm{sample}}$ and phase difference $\delta\phi = \pi + \phi_\mathrm{ref}-\phi_\mathrm{sample}$ at equilibrium: $C_\mathrm{sample}= \cos(\delta\phi) R C_\mathrm{ref}   $.

\subsection{Design and fabrication details}

As discussed earlier, several different designs and fabrication recipes were used throughout this work to fabricate devices. 
In the first part, we give some general information on steps that have been employed for many of these fabrication runs. Next we describe fabrication processes of different dielectrics used in the first design to separate the top and grid gate layers.
Finally, we provide detailed information for a fabrication run of the second design with overlapping aluminum gates, which serves as a clear example from which the steps required for fabricating the other devices measured can be  deduced.\\
All lithography steps were performed using electron beam lithography (Vistec EBPG 5000+ or 5200) at 100kV acceleration voltage. Etching Al$_{x}$Ga$_{1-x}$As was done using diluted Piranha (1:8:240 H$_2$SO$_4$:H$_2$O$_2$:H$_2$O) yielding etch rates of roughly 4 nm/s. The actual etch rate decreases on a timescale of minutes as the H$_2$O$_2$ concentration slowly decreases.
Spinning is done at 500 rpm for 5 seconds and then for 55 seconds at speeds listed below.
Etching SiO$_2$ and AlO$_x$ was done using buffered HF (BOE 1:10) solutions. After either type of wet etch, devices are rinsed repeatedly in H$_2$O. Adhesion issues for resists with HF etch times longer than 20 s mean iterative etching and re-baking is necessary. For the 366 nm AlO$_x$ layers, this meant we had to use a dry Cl etch to etch the bulk of the depth of the vias before finishing with a wet etch.
Metallic layers were deposited using electron-beam evaporation at room temperature and subsequent lift-off in a solvent.\\
The first design, with a thick dielectric separating the two gate layers, has been fabricated with two different dielectrics.
For the results of Fig. 5a-b in the main text, plasma-enhanced chemical vapor deposition (PECVD) of 360 nm of SiO$_2$ as dielectric was used, which was found to introduce phase-noise during capacitance-bridge measurements. We have also fabricated devices with 366 nm of plasma-enhanced atomic layer deposition (ALD) grown AlO$_x$ dielectric (optical image in Fig. S 1a).
Although these devices had less phase-noise, they showed large top gate hysteresis, rendering them practically impossible to measure with (Fig. S 1b). Furthermore, etching small vias through such a thick layer of alumina is very cumbersome. All devices of the first design had low yield in wire-bonding because of poor adhesion of the dielectric layers to the GaAs surface.

An overview of the fabrication steps for realizing double-layer gate devices with aluminum gates (second design) is given below. See Fig. S 2a for schematic
side views of the process and Fig. S 2e for the top view of a finished device.

\begin{figure}[!htb]
	\centering
	\includegraphics{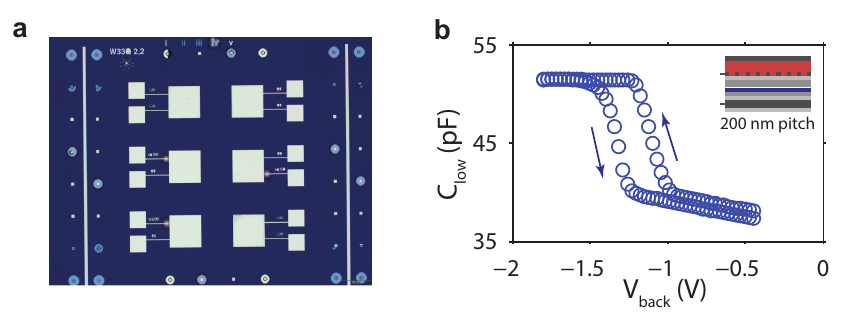}
	\caption{(a) Optical image of a device cell of the first design with a 366 nm thick AlO$_x$ dielectric layer. Top left and bottom right squares are ohmic contacts, which could have also been fabricated on the back side of the wafer. The other squares are three double-gate devices and one single-gate device, each with two bond pads. Contacting the grid layer underneath the dielectric is done using etched vias.
	(b) Capacitance as a function of back gate for a device from (a), showing hysteresis as function of either top or back gate voltage (shown).}
	\label{fig:figS3}
\end{figure}

\begin{figure}[!htb]
	\centering
	\includegraphics{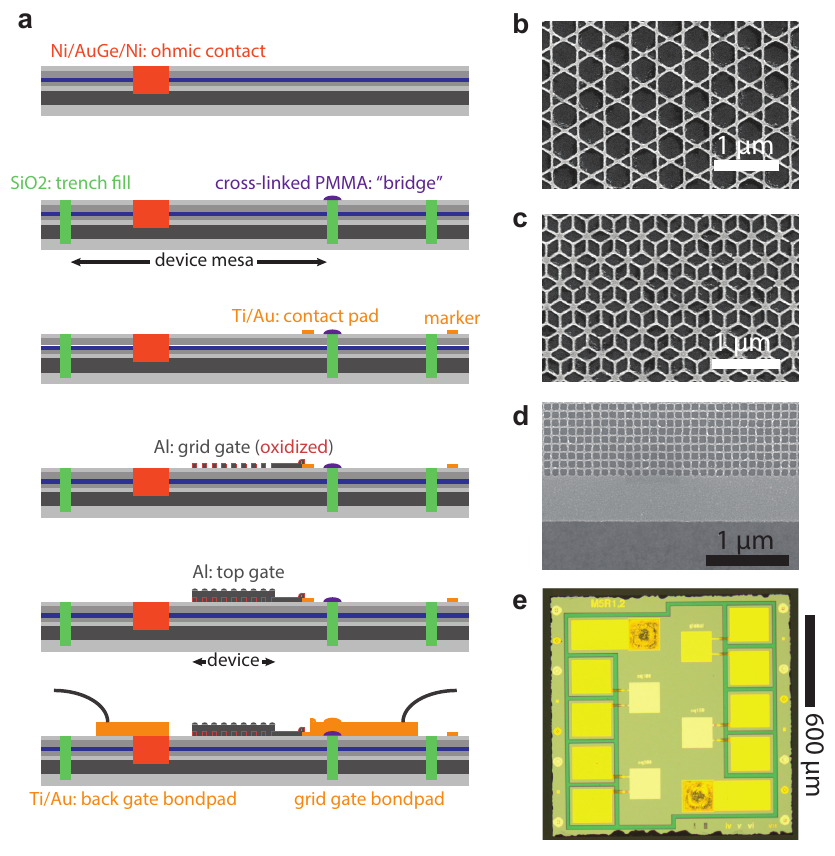}
	\caption{(a) Overview of the fabrication process for overlapping gates. Note that the top gate bondpad is not shown in this view, and that the mesa and bridge steps are shown added to the same figure. (b) Electron micrograph of a kagome-type lattice fabricated using the same lift-off recipe as used for the square grids, with 5/15 nm Ti/Au. Note that for non-square lattice such as this, a different lattice is imposed on the 2DEG depending on the polarity of the voltage difference between the grid and top gate.
    (c) Electron micrograph of a 5/15 nm Ti/Au dice-type lattice gate.
    (d) Electron micrograph at the edge of a square 100 nm pitch grid, showing the overdosed 'frame' that is used to counter proximity effect induced inhomogeneity effects at the edge of the grid.
    (e) Optical image of a device cell containing two ohmic contacts at top left and bottom right as well as a single-gate device on the top right and three double-gate devices. The dielectric-filled etched region that separates the device mesa from the bonding pads is visible in green.
    }
	\label{fig:fab}
\end{figure}

\begin{itemize}
\item Ohmic contacts - spin PMMA 495K A8 resist at 6000 rpm - bake 15 min at 175 $^\circ$C (400 nm) - lithography - development 60 s in 1:3 MIBK/IPA - wet etch of 180 nm in diluted Piranha - evaporation of 5/150/25 nm Ni/AuGe/Ni - lift-off in acetone and IPA rinse - anneal 60 s at 440 $^\circ$C in forming gas.
\item Mesa etch - spin PMMA 495K A8 resist at 6000 rpm - bake 15 min at 175 $^\circ$C (400 nm) - lithography - development 60 s in 1:3 MIBK/IPA - wet etch of 700 nm in diluted Piranha - sputtering 700 nm of SiO$_2$ - lift-off in acetone and IPA rinse.
\item Bridges - spin PMMA 495K A8 resist at 6000 rpm - bake 15 min at 175 $^\circ$C (400 nm) - lithography - cross-link PMMA strips through electron beam overdose at $25$ mC/cm$^2$. These sections act as bridges over which the leads will connect sample mesa and bond pad regions.
\item Connection pads and markers - spin PMMA 495K A8 resist at 6000 rpm - bake 15 min at 175 $^\circ$C (400 nm) - lithography - development 60 s in 1:3 MIBK/IPA - evaporation of 10/50 nm Ti/Au - lift-off in acetone and IPA rinse. These sections act either as markers or as pads that will be contacted on the top both by the Al gates and the leads contacting the bond pads. We found these thin layers of metal to be the most robust way to make an electrical connection (typically several Ohm) between the Al gates and the Au bond pads.
\item Grid gate - spin CSAR 62.04 resist at 5000 rpm - bake 3 min at 150 $^\circ$C (72 nm) - lithography - development 70 s pentyl acetate and 60 s 1:1 MIBK:IPA - evaporation of 20 nm Al - lift-off in NMP at 70 $^\circ$C using soft ultrasound excitation for 4 hrs and subsequent acetone and IPA rinse - oxidation in 20 min at 200 $^\circ$C at 100 mTorr and 300 W RF power using the remote plasma of an ALD machine. We have optimized the lithographic sequential writing such that a 200 $\upmu$m x 200 $\upmu$m grid is written in one go and at under a minute, avoiding stitching errors and reducing the effect of drift (typically several tens of nm/min). We have done this by direct programming of an iterative sequence that the e-beam follows in writing the grid instead of the standard procedure of converting a design file (in this case a large square grid) to an e-beam lithography file using BEAMER software. Furthermore, we add a 200 nm thin frame around the grids whose overdose is chosen to counter proximity edge effects (Fig. S 2d). Note also that we found the conflicting requirements of high resolution and undercut required for lift-off to be best served using a single layer CSAR62 resist. Finally, we find feature size, yield and reproducibility to be limited by the grain size of the evaporated Al, instead of by the resist mask or lithography process. To achieve a smaller grain size, we used a fast Al evaporation rate of 0.2 nm/s. As such, Ti/Au but especially Ti/AuPd gates are easier to fabricate than Al gates but they cannot be oxidized and would require actual deposition of a dielectric. Also note that the lift-off based fabrication of grids allows for different lattice types to be made, see Fig. S 2b-d.
\item Top gate - spin PMMA 495K A8 resist at 6000 rpm - bake 15 min at 175 $^\circ$C (400 nm) - lithography - development 60 s in 1:3 MIBK/IPA - evaporation of 50 nm Al - lift-off in acetone and IPA rinse.
\item Bonding pads - spin OEBR-1000 (200cp) lift-off resist at 3500nm - bake 30 min at 175 $^\circ$C (500 nm) - spin PMMA 950K A2 resist at 2000 rpm - bake 10 min at 175 $^\circ$C (90 nm) - lithography - evaporation of 50/200 nm Ti/Au -  lift-off in acetone and IPA rinse.
\end{itemize}

\subsection{Conversion from capacitance to density of states}

In calculating density of states from capacitance data, we follow a procedure described before\cite{Ashoori1991}.
We model the system as a parallel plate capacitor made up of the top and bottom gates, with the potential for added charges at the location of the quantum well, as sketched in Fig. S 3.
As a start, $C_\textup{high/low}$ are measured as function of gate voltages and magnetic field values (Fig. S 4a).
Note that the heterostructure stack is designed to keep the tunnel frequency in the middle of the experimental measurement window (Fig. 4a-b in the main text): below 1 kHz signal to noise ratio declines (mainly because of the $1/f$ noise of the first transistor in the amplification chain) and above 2 MHz systematic errors occur (we find asymmetric cross-talk between the two excitation signals and the second transistor in the amplification chain).

The total voltage difference over the device is a combination of the electric fields $V=V_\mathrm{back}-V_\mathrm{top}=E_1(w+d)+E_2d$, which in turn depends on the charges on the plates as $V=\frac{\sigma_\mathrm{top}(w+d)}{\epsilon} + \frac{\sigma_\mathrm{QW}d}{\epsilon}$.
The total capacitance, which is the one measured at low enough frequencies, is defined as $C_\mathrm{low} = \frac{\partial Q}{\partial V} = A\frac{\partial \sigma_\mathrm{top}}{\partial V} = \frac{\epsilon A}{w+d} - \frac{dA}{w+d} \frac{\partial \sigma_\mathrm{QW}}{\partial V} + $ small terms that depend on changing distances and which we ignore.
The first term describes the bare capacitor, and is therefore equal to the total capacitance measured at high frequencies: $C_\mathrm{high}= \frac{\epsilon A}{w+d}$. The second term is the one of interest.
It describes changes between the capacitance measured at low and high frequency because of the addition of charges in the quantum well, which allows us to infer changes in electron density using $\frac{\partial n}{\partial V}=-\frac{1}{e}\frac{\partial \sigma_\mathrm{QW}}{\partial V}=\frac{1}{eA} \frac{w+d}{d} \left( C_\mathrm{low}-C_\mathrm{high} \right)$ (Fig. S 4b).\\

\begin{figure}[!htb]
	\centering
	\includegraphics{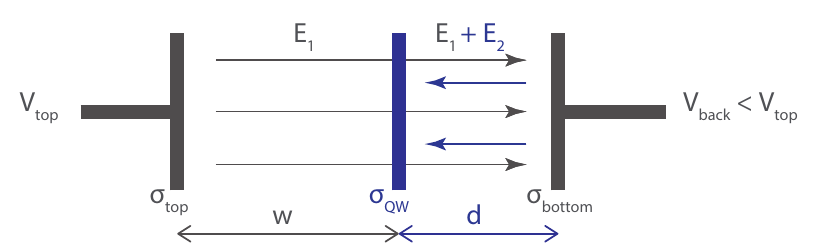}
	\caption{Schematic representation of the device as a parallel place capacitor of distance $w+d$ with an inserted quantum well at a distance $d$ from the back gate. When the DOS at the quantum well is nonzero, charges can build up.}
	\label{fig:figS1}
\end{figure}

The voltage required to change the Fermi level $E_\mathrm{F}$ of the quantum well can be found using a similar deduction to the one described above, and is described by a voltage-dependent lever arm $\alpha \equiv -e\frac{\partial V}{\partial E_\mathrm{F}}$.
We find the lever arm by following the dependence of the Fermi level in the quantum well through changes in the electric field as $\frac{\partial E_\mathrm{F}}{\partial V} = -e  w \frac{\partial E_1}{\partial V} = -e \left( \frac{w}{w+d} + \frac{e}{\epsilon}\frac{wd}{w+d} \frac{\partial n}{\partial V} \right)$ (Fig. S 4c).
The first term describes how the Fermi level of a gapped system in the quantum well ($\delta n=0$) changes with bias as expected given its relative location $\frac{w}{w+d}$ between the plates of a simple parallel plate capacitor (Fig. S 4c).
It is the second term that encompasses the electron filling, showing the lever arm to increase when charges can be added to the quantum well (after accumulation this becomes the dominant term, see Fig. S 4b).
Given the above expressions for density and Fermi level changes as function of gate voltage, we can define the density of states in the 2DEG through $DOS = \frac{\partial n}{\partial V} \frac{\partial V}{\partial E_\mathrm{F}} = \frac{1}{e^2 A} \frac{w+d}{d} \alpha \left( C_\mathrm{low}-C_\mathrm{high} \right)$ (Fig. S 4d).
\begin{figure}[!htb]
	\centering
	\includegraphics{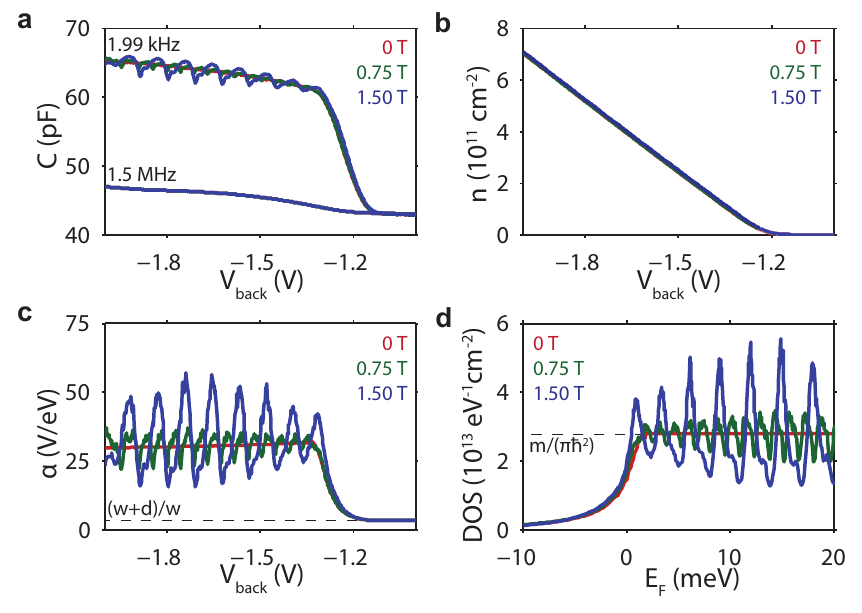}
	\caption{(a) Capacitance as function of back gate bias at 0 T (red), 0.75 T (green) and 1.5 T (blue), both below and above the tunnel frequency. Top gate is kept constant for all measurements as it is serves as the bridge point, which is also directly contacted to the gate of the first transistor in the amplification chain.
	(b) Density as function of back gate bias. As the system becomes more gapped between Landau levels at higher fields, steps start to form in the graph that indicate the filling of distinct levels at well-defined densities.
	(c) Lever arm as function of back gate bias. Note that the quantum capacitance of a large density of states in the 2DEG increases the voltage required to change the Fermi level, as expected. At zero density of states, however, the lever arm is simply the geometric ratio expected from the relative location of the quantum well between the top and bottom gate.
	(d) Density of states as function of the Fermi energy, which is the integrated lever arm. We choose zero in energy to lie close to accumulation.}
	\label{fig:DOS}
\end{figure}
As indicated by changes in $C_\mathrm{high}$ in Fig. S 4a, the distances describing the system are non-static with gate voltage. In the case of $(w+d)$, this is most likely due to back gate charges populating part of the spacer layer as the electric fields bend the conduction band edge, indeed increasing $C_\mathrm{high}$ for more negative back gate voltages.
The exact location of the charges in the quantum well and related distance $d$, however, we cannot directly infer from an independent measurement.
As a first guess, the growth distances combined with the $(w+d)$ extracted from $C_\mathrm{high}$ suffices.
A better estimate can be made using the known linear degeneracy of Landau levels with magnetic field, $n_\mathrm{LL} = \frac{2eB}{h}$ (Fig. S 4b).
To obtain the best possible calibration, however, we compensate for any further dependence of the relative quantum well position on back gate voltage by pegging the 0 T DOS after accumulation to the expected value of $\frac{m}{\pi \hbar^2} \approx 2.8 \times 10^{13}$ eV$^{-1}$cm$^{-2}$ (Fig. S 4d), and use this calibration for nonzero magnetic field values.

\end{document}